\documentclass{PoS}

\title{Centrifugal acceleration of protons in the vicinity of a supermassive black hole}

\ShortTitle{Acceleration of protons in the AGN}

\author{\speaker{Gunya A. A.}\\
	The P.N. Lebedev Physical Institute, LPI
	E-mail: \email{aagunya@lebedev.ru}}

\author{Istomin Y. N.\\
	The P.N. Lebedev Physical Institute, LPI
	E-mail: \email{istomin@td.lpi.ru}}

\abstract{	Acceleration of protons in the active galactic nuclei is considered.
	The largest energy is achieved by protons during centrifugal acceleration in the magnetosphere of the central machine. When the charged test particle accelerated in the magnetosphere of a black hole approaches light cylinder surface, acceleration occurs mainly in the azimuthal direction, i.e. the acceleration is centrifugal. In this paper the acceleration of a proton having smaller synchrotron losses compared to the electron is considered. As a proton experiences the highest energy increase while accelerating near the light surface, a partial solution for the maximum Lorentz factor can be obtained there. In the analysis the obtained dependence of the maximum energy on the parameter of particle magnetization $ \kappa $ and parameter $ \alpha $ which reflects the relation of toroidal $ B_\phi $ and poloidal $ B_T $ magnetic fields , has led to the conclusion that the achievement of theoretical maximum limit of Lorentz factor value $ \gamma_m=\kappa^{-1}$ is not possible for an accelerated particle in the magnetosphere of a black hole due to restrictions of the topology of toroidal and poloidal magnetic fields imposed. The analysis of special cases of the relation of toroidal and poloidal magnetic field has shown that in the presence of magnetic field that is significantly more toroidal the maximum Lorentz factor value reaches $\gamma_m = \kappa^ {-2/3} $, in case when toroidal field becomes smaller in comparison to poloidal field the maximum Lorentz factor value does not exceed $\gamma_m = \kappa^ {-1/2} $. 
	
	For a number of objects, such as M87 and Sgr. A *, maximum Lorentz factor values for accelerated protons for scenarios of existence or lack of toroidal magnetic field have been derived. The obtained results for magnetosphere of Sgr. A *  has confirmed by the experimental data obtained on the massive HESS of Cherenkov telescopes.}

\FullConference{High Energy Phenomena in Relativistic Outflows VII - HEPRO VII\\
	9-12 July 2019\\
	Facultat de Física, Universitat de Barcelona, Spain}

\begin{document}
	
	\section{Introduction}
	
	For the last decades Cherenkov telescopes (such as VERITAS, HESS, MAGIC, etc.) have detected gamma-ray sources of multi-TeV energies. Many of these sources are active galactic nuclei. The candidates can be labelled the "laboratory" galaxy of M87, 3C279, NGC 5128, etc.  In addition, the analysis of the VHE emission from the Sgr A* region suggests that the radiating particles (protons) need to have energy in excess of $ 10^15 $ eV (Abramowski et al. 2016).
	As majority of carriers, the protons with energies $\simeq 10^{20}$ eV act similarly to electrons with energies in the TeV range.
	To explain the origin of suchlike particles the mechanism of centrifugal acceleration was proposed by Rieger and Manheim (2000). As a result of synchrotron radiation, electrons lose a considerable part of energy, and their acceleration is not as effective as acceleration of protons, being a main cosmic-ray component (Ginzburg 1966). Acceleration of charged particles in a magnetosphere is bound to the presence of an electric field proportional to the rotation angular velocity of the magnetospheric plasma involved in rotation with a black hole (Blandford and Znaek 1977), and an accretion disk (Blandford and Payne 1982). Acceleration of protons in a disk can be considered as an initial process of pre-acceleration of the particles injected from a disk in a magnetosphere. However, in view of proton interactions with the photon field of a disk (disk temperature about 10 eV), maximal Lorentz factor is unable to reach larger values (Istomin and Sol, 2009).
	The structure of the magnetic field is described in section 2. Equations of motion of charged particles are presented in section 3, where the approximate analytical expression for the maximum energies is also derived. The obtained results are discussed in section 4 in the context of protons acceleration in a number of real AGN.
	
	\section{Fields structure}
	
	For split monopole, magnetic field lines come out directly from the vicinity of the central black hole. They are involved into the rotation by the black hole and rotate with an angular velocity of $ \Omega_F $, which for optimal matching is half the angular velocity of rotation of the black hole
	$ \Omega_H $, $ \Omega_F = \Omega_H / 2 $ (Blandford and Znajek, 1977). Electric field must be $ E_\theta = -u_ \phi B_r / c = -r \Omega_F B_r \sin \theta / c $. The magnetosphere consists of electromagnetic components:
	$$
	B_r=\sigma B_0\left(\frac{r}{r_L}\right)^{-2},
	$$
	\begin{equation}\label{fields}
		B_\phi=\alpha\sigma B_0\left(\frac{r}{r_L}\right)^{-1}, \,
	\end{equation}
	$$
	E_\theta=-\sigma B_0\sin\theta\left(\frac{r}{r_L}\right)^{-1}.
	$$	
	
	Here, $ r_L = c / \Omega_F $ is the natural size of the magnetosphere in the transverse direction.
	$ \alpha $ parameter of the ratio of the toroidal magnetic field $ B_T $ to the poloidal one $ B_P $.
	$ \sigma $ is the sign of $ z $ , $ \sigma = sign(z) $.
	
	\begin{figure}
		\centering{\includegraphics[width=1\linewidth]{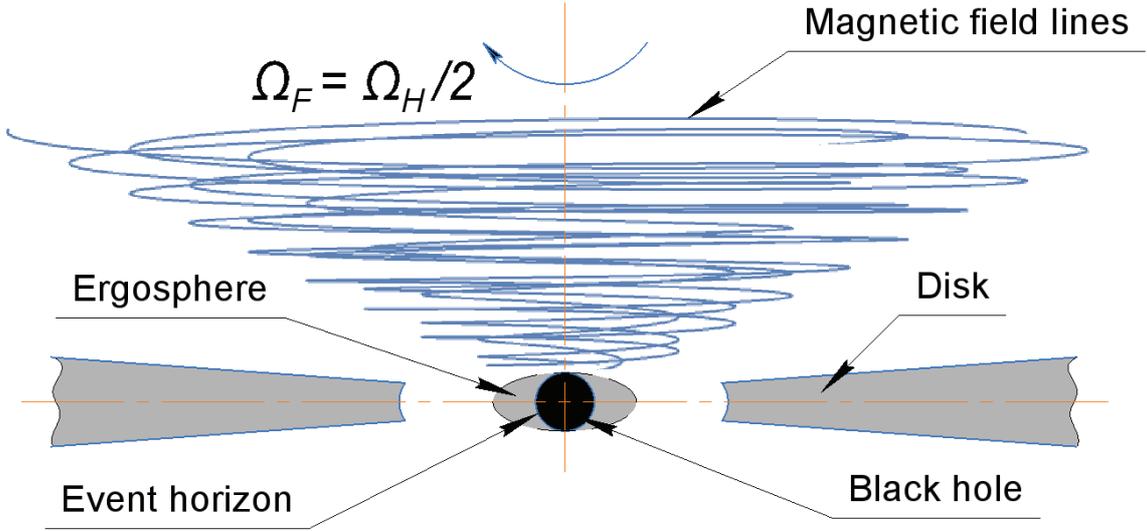}}
		\caption{Schematic picture of magnetic field lines in the magnetosphere in the vicinity of the rotating massive black hole}
		\label{ris:image1}
	\end{figure}
	
	The configuration of magnetic field lines in the magnetosphere, defined by the ratio of $dr / B_r = rd \phi / B_\phi$, yields spirals $r = r_1 \phi + r_0$, where the values of $r_0$ and $r_1$ are constants. Figure 1 shows the magnetic field lines of this configuration in the magnetosphere.
	
	\section{Acceleration} 
	
	The main equations of particle motion include the momentum $ \bf p $ and coordinate $ \bf r $ :
	
	$$\frac{d{\bf p}}{dt}=q\left({\bf E}+\frac{1}{c}\left[{\bf v,B}\right]\right), \,$$
	\begin{equation}\label{m}
		\frac{d{\bf r}}{dt}={\bf v}=\frac{{\bf p}}{m\gamma}, \, 
	\end{equation}
	
	Here, $ \gamma $ is Lorentz factor.
	
	$$\gamma^2=1+\frac{p^2}{m^2 c^2}.$$
	
	To further develop numerical equations, let us introduce dimensionless time, coordinates, velocity and momentum,
	
	\begin{equation}\label{dimensionless}
		t'=\frac{\omega_c t}{\gamma_i}, \, 
		{\bf r'}=\frac{{\bf r}}{r_L}, \, {\bf v'}=\frac{{\bf v}}{c}, \, 
		{\bf p'}=\frac{{\bf p}}{mc\gamma_i}.
	\end{equation}
	
	Here, $ \omega_c $ is non relativistic cyclotron frequency, $ \gamma_i $ is initial value of Lorenz factor. 
	
	Taking into account the field equations (\ref{fields}), we get dimensionless equations in the spherical coordinates $ \bf r $, $ \bf \theta $, $ \bf \phi $
	
	\begin{eqnarray}\label{m1}
		&&\frac{dp_r}{dt}=\frac{\kappa}{r\gamma}\left(p_\theta^2+p_\phi^2\right)+\frac{\sigma\alpha}{r\gamma}p_\theta, \nonumber \\ 
		&&\frac{dp_\theta}{dt}=-\frac{\kappa}{r\gamma}\left(p_r p_\theta-p_\phi^2 \cot\theta\right)-\frac{\sigma}{r}\sin\theta+\frac{\sigma}{r^2\gamma}p_\phi -\frac{\sigma\alpha}{r\gamma}p_r, \nonumber \\  
		&&\frac{dp_\phi}{dt}=-\frac{\kappa}{r\gamma}\left(p_r+p_\theta \cot\theta\right)p_\phi- \frac{\sigma}{r^2\gamma}p_\theta, \\  
		&&\frac{dr}{dt}=\frac{\kappa}{\gamma} p_r, \nonumber \\  
		&&\frac{d\theta}{dt}=\frac{\kappa}{r\gamma}p_\theta.  \nonumber
	\end{eqnarray}
	
	Here, $ \alpha $ and $ \kappa $ are dimensionless parameters. The $ \alpha $ value is equal to the ratio of the toroidal magnetic field to the poloidal one at the distance $ r = r_L $, $ \alpha = B_ \phi / B_r |_{r = r_L} $. The value of $ \kappa $ is the parameter of particles magnetization (or freeze-in parameter), it reflects the relationship of magnetic field with particle and equals to
	
	\begin{equation}\label{kap}
		\kappa=\frac{c\gamma_i}{r_L\omega_c}=\frac{r_c}{r_L}=\frac{\Omega_F}{\omega_c/\gamma_i}
	\end{equation}
	
	Here, $ r_c $ is cyclotron radius. The first important dependence (Figure 2) we get from system \ref {m1} is Lorenz factor versus the radial distance to light cylinder, where the particle achieves maximum value of Lorenz factor $ \gamma_m $.
	
	\begin{figure}
		\centering{\includegraphics[width=1\linewidth]{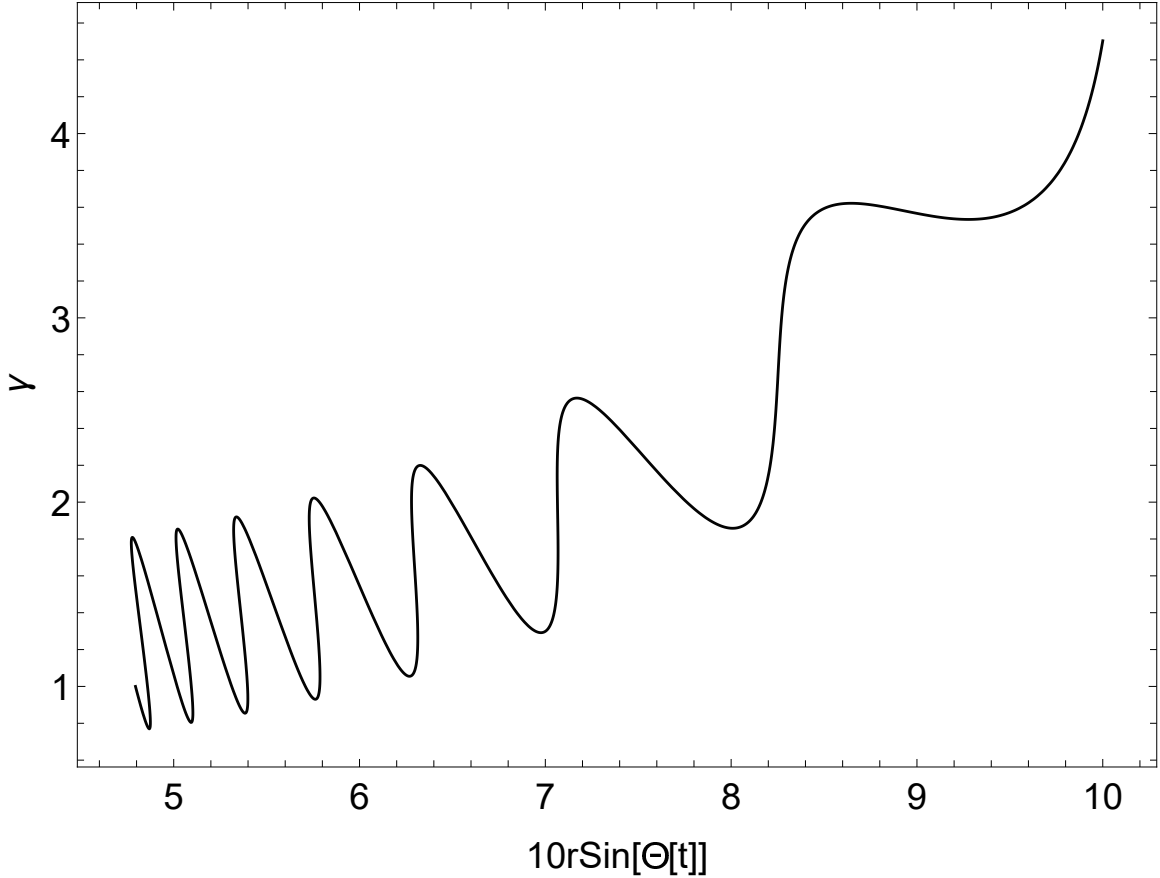}}
		\caption{The Lorentz factor of the particle $ \gamma_m $ versus the radial distance $ r\sin\theta $ for $ \kappa = 10^{-2} $ and $ \alpha = 10^{-2} $. The light cylinder surface is located at 10 on the abscissa axis}
		\label{ris:image2}
	\end{figure}
	
	Since the main acceleration occurs near the light surface (Figure 2), then $ \Delta \theta = \theta - \theta_0 = - \sigma \kappa
	\gamma_m / \sin \theta $. The angle $ \theta $ varies strongly only in the vicinity of the light surface, practically without changing the main part of the trajectory. On the other hand, $ d \theta / dr = p_\theta \sin\theta / p_r = \Delta \theta / \Delta r $,
	where $ \Delta r $ is the size of the region near the light surface $ r = 1 / \sin \theta $, where the main acceleration of particles occurs
	
	\begin{equation}\label{delta}
		\Delta r=-\frac{\sigma\kappa\gamma_m}{\sin^2\theta}\frac{p_r}{p_\theta}\mid_{r=1/\sin\theta}.
	\end{equation}
	
	Passing in the system (\ref {m1}) from time derivatives to derivatives over coordinate $r, \, d/dt=d/dr(p_r/\gamma)$. Replacing derivatives by fractions $d{\bf p}/dr\simeq {\bf p}|_{r=1/\sin\theta}/\Delta r$, we obtain an algebraic system of equations that determines the values of ($\gamma_m, {\bf p})|_{r=1/\sin\theta}$ at the light surface $r=1/\sin\theta$
	
	\begin{eqnarray}\label{algebra}
		&&p_r^2+\frac{\alpha\gamma_m}{\sin\theta}p_r-\frac{\gamma_m(\sin^2\theta+\kappa\gamma_m^2|\cos\theta|)}{\sin^2\theta}=0; \nonumber \\
		&&p_\theta =-p_r\frac{\sigma\kappa\gamma_m^2\sin\theta}{\sin^2\theta+\kappa\gamma_m^2|\cos\theta|}; \\
		&&p_\theta^2+p_r^2=2\gamma_m.  \nonumber 
	\end{eqnarray}
	
	Introducing $ \alpha = 0 $ in the system of equations (\ref{algebra}) we get
	
	\begin{equation}\label{alpha0}
		\gamma_m=\kappa^{-1/2}\sin\theta.
	\end{equation}
	
	This result coincides with the expression obtained in Istomin and Sol (2009) for the case of $ \theta = \pi / 2 $, where the acceleration of particles in the magnetosphere was considered only near the accretion disk. The expression $ \gamma_m = \kappa^{-1/2} (\sin \theta = 1) $ was obtained from the analysis of numerical calculations of particle trajectories, while here we use the analytical approximation.
	
	\begin{figure}
		\centering{\includegraphics[width=1\linewidth]{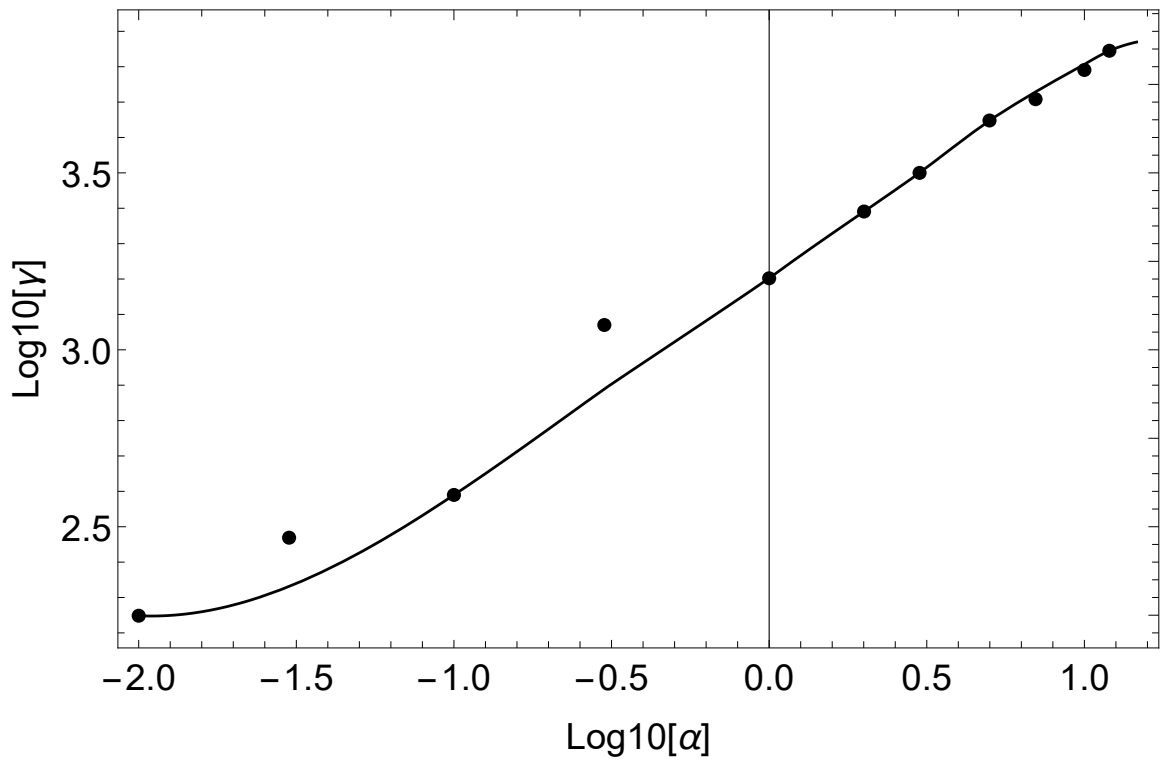}}
		\caption{Lorentz factor of proton $ \gamma $ versus $ \alpha $ for $ \kappa = 10^{-4} $}
		\label{ris:image3}
	\end{figure}
	
	To obtain a general expression for $ \gamma_m $ for the arbitrary value of $ \alpha $, one needs to solve a quadratic equation, the first equation of the system (\ref{algebra}), with respect to the value of $ p_r $. However, it is preferable to find an approximate expression for the value of $ p_r $ for the case of large values of $ \alpha, \, \alpha^2 \gamma_m >> 1, $
	
	\begin{equation}\label{pr}	
		p_r\simeq \frac{\sin^2\theta+\kappa\gamma_m^2|\cos\theta|}{\alpha\sin\theta}. 
	\end{equation}
	
	The result is
	
	\begin{equation}\label{alpha}
		\gamma_m\simeq 2^{1/3}\left(\alpha\kappa^{-1}\right)^{2/3}\sin^{2/3}\theta.
	\end{equation}
	
	Figure 3 depicts the dependence of $ \gamma_m $ on the toroidal magnetic field magnitude.	
	The dots correspond the values calculated numerically using the equations of motion of particles in the magnetosphere (\ref {m1}). Clearly visible is the transition from being independent $ \alpha $ at small values of $ \alpha $ to the relation $ \gamma_m \propto \alpha ^ {2/3} $
	
	\begin{figure}
		\centering{\includegraphics[width=1\linewidth]{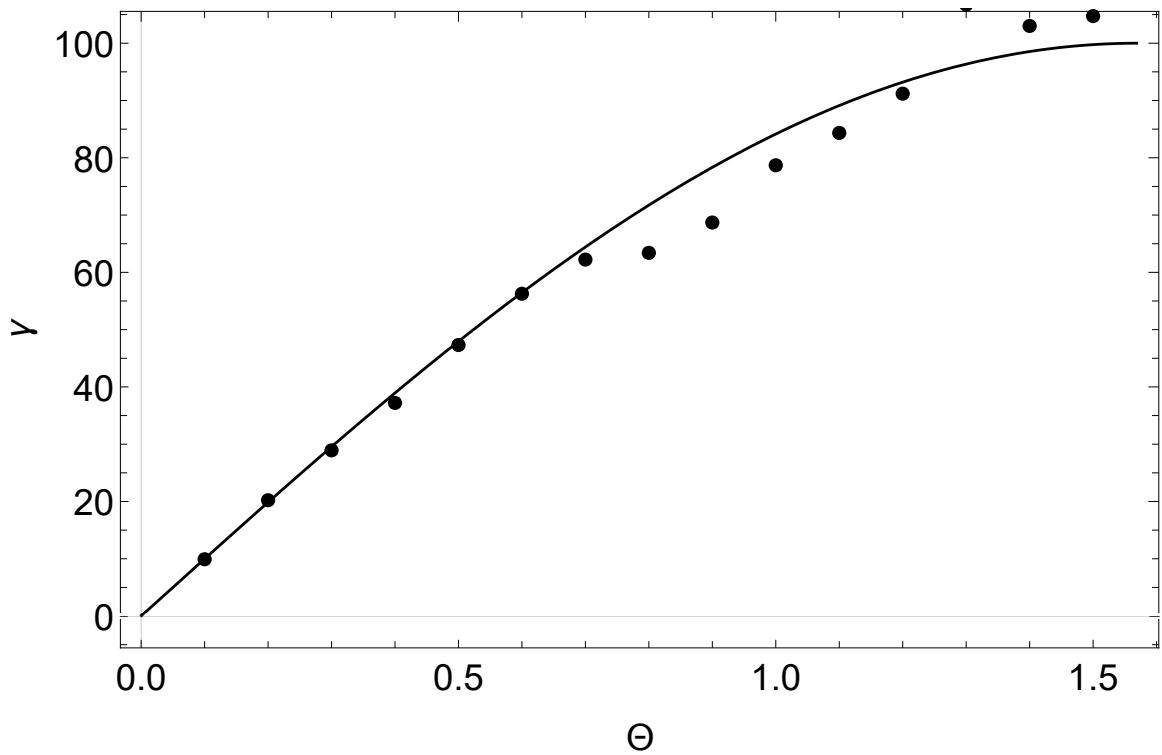}}
		\caption{Lorentz factor of a particle at light surface $ \gamma_m $ versus the polar angle $ \theta $ for $ \kappa = 10^{-4} $}.
		\label{ris:image4}
	\end{figure}
	
	The dependence (Figure \ref {ris:image4}) of $ \gamma_m $ on the angle $ \theta $ means that at different heights $ z $ above the equatorial plane the particle energy is also different.
	Accelerated particles, leaving the vicinity of a black hole, will form a single energy spectrum.
	In summary, crossing the light surface $ r = r_L / \sin \theta $ and acquiring energy $ \gamma_m $, depending on the angle $ \theta $, we can obtain the distribution function of fast particles $ f (\gamma_m) $. Because $ z = r_L \cot \theta, \, dz = -d \theta / \sin^2 \theta $, and $ f (\gamma_m) d \gamma_m \propto \pi r_L^2 ndz $, where $ n $ is the density of the 'primary' particles in the magnetosphere inside the light surface ($ n \simeq const $), we get
	
	\begin{eqnarray}\label{function}
		&&f(\gamma_m)\propto\gamma_m^{-2}, \, \gamma_m<\gamma_0^{1/2}, \,  \alpha<\gamma_0^{-1/4}; \nonumber  \\
		&&f(\gamma_m)\propto\gamma_m^{-2.5}, \, \gamma_m<\gamma_0^{2/3}, \,  \alpha>\gamma_0^{-1/4}.
	\end{eqnarray}
	
	Here we assume that angles $ \theta $ really change from some small value of $\theta_1 $, determined by the height of the magnetosphere, to $ \theta_2 <\pi / 2 $. The angle $ \theta_2 $ separates the region of the magnetosphere adjacent to the accretion disk, where the matter rotates with the speed close to the rotation speed of the disk, and the region where the matter rotates with the speed close to the rotation speed of the black hole. 
	
	Figures \ref {ris:image3} and \ref {ris:image4} show the comparison of numerical calculations (dots) with the analytical expressions. The dots of the numerical calculations in figures \ref {ris:image3} and \ref {ris:image4} were obtained from \ref {m1} by the method of finite number of variation $ \alpha $ parameter with $ \kappa = const $.
	
	\section{Results} 
	
	\begin{table}[htb]
		\begin{center}
			\begin{tabular}{|c|c|c|c|}
				\hline
				$ \gamma_m $ & $\kappa$ & $E_{max}$ & Case \\
				\hline
				$ \gamma_0 $ & $ \kappa ^{-1} $ & $ 3.8 \cdot 10^{21} M_9 B_4 $ & Theoretical $E_{max}$ is not achieved in the magnetosphere \\
				$ \gamma_0 ^ {2/3}$ & $ \kappa ^{-2/3} $ & $ 2.3 \cdot 10^{17} (M_9B_4)^{2/3} $ & Split monopole plus a significant toroidal field  \\
				$ \gamma_0 ^ {1/2}$ & $ \kappa ^{-1/2} $ & $ 1.9\cdot 10^{15}(M_9B_4)^{1/2} $ & Weak toroidal field \\
				\hline
			\end{tabular}
		\end{center}
		\caption{\label{tab:cases} Cases of maximum energy achieved depending on magnetic field configuration}
	\end{table}
	
	Here the value of $ M_9 $ is the mass of a black hole measured in units of $ 10^9$, and the value of $ B_4 $ is the magnitude of the magnetic field in units of $ 10^4 \, G $.
	Estimates of $\gamma_m$ are Lorentz factor achieved by a 'cold' particle, $\gamma_i \simeq 1$, as a result of its acceleration in the magnetosphere. If the particle has undergone preliminary acceleration, $\gamma_i > 1$, then the maximum possible energies increase substantially. The results obtained here can also be used for microquasars. The $E_{max, 1/2}$ for Sgt. A* presented in Table 2 is consistent with the observations of the Galactic centre by the HESS Cherenkov telescope (Abramowski et al., 2016).
	
	\begin{table}[htb]
		\begin{center}
			\begin{tabular}{|c|c|c|c|c|c|}
				\hline
				Galaxy & $M_9$ & $B_4$ & $E_{max, 1} $ & $E_{max, 1/2} $ & $E_{max, 2/3} $\\
				\hline
				Sgt. A* & 0.0043 & 0.36 & $5.8\cdot10^{18}$ & $7.4\cdot10^{13}$ & $ 3.2\cdot10^{15} $ \\
				NGC4486 & 6.6 & 0.45 & $8.9\cdot10^{21}$ & $2.9\cdot10^{15}$ & $4.2\cdot10^{17}$\\
				NGC1399 & 5.2 & 0.44 &$9.2\cdot10^{21}$ &  $2.9\cdot10^{15}$ & $3.1\cdot10^{17}$\\
				NGC4874 & 20.8 & 0.09 &$7.1\cdot10^{21}$ &  $3.6\cdot10^{15}$ & $4.3\cdot10^{17}$\\
				NGC4889 & 26.9 & 0.06 &$6.1\cdot10^{21}$ &  $3.1\cdot10^{15}$ & $3.7\cdot10^{17}$\\
				NGC6166 & 28.4 & 0.06 &$6.5\cdot10^{21}$ &  $3.2\cdot10^{15}$ & $3.9\cdot10^{17}$\\
				\hline
			\end{tabular}
		\end{center}
		\caption{\label{tab:galaxyes}Upper limits for proton acceleration. The proton energies are given in units eV}
	\end{table}
	
	The data presented in the works of Maggorian et al. (1998), as well as Boldt and Ghosh (1999) and Broderick et al. (2015), contain magnetic field and black holes masses values for some of the most famous AGN.

\end{document}